\documentclass{article}

\PassOptionsToPackage{numbers, compress}{natbib}

\usepackage[preprint]{neurips_2022}

\usepackage[utf8]{inputenc} % allow utf-8 input
\usepackage[T1]{fontenc}    % use 8-bit T1 fonts
\usepackage{hyperref}       % hyperlinks
\usepackage{url}            % simple URL typesetting
\usepackage{booktabs}       % professional-quality tables
\usepackage{amsfonts}       % blackboard math symbols
\usepackage{nicefrac}       % compact symbols for 1/2, etc.
\usepackage{microtype}      % microtypography
\usepackage{xcolor}         % colors
\usepackage{wrapfig}
\usepackage{amsfonts}
\usepackage{amsmath}
\usepackage{graphicx}
\usepackage{algorithm}
\usepackage{algpseudocode}

\title{Neural Sound Field Decomposition with Super-resolution of Sound Direction}

\author{%
  Qiuqiang Kong$^{1}$, Shilei Liu$^{1}$, Junjie Shi$^{1}$, Xuzhou Ye$^{1}$ \\ \textbf{Yin Cao$^{2}$, Qiaoxi Zhu$^{3}$, Yong Xu$^{4}$, Yuxuan Wang$^{1}$} \\
  $^{1}$ ByteDance, Shanghai, China \\
  $^{2}$ University of Surrey, Guildford, UK \\
  $^{3}$ University of Technology Sydney, Sydney, Australia \\
  $^{4}$ Tencent AI Lab, Bellevue, USA \\
  $^{1}$\texttt{\{kongqiuqiang, liushilei.666, shijunjie, yexuzhou, 
wangyuxuan.11\}@bytedance.com} \\
  $^{2}$yin.cao@surrey.ac.uk, $^{3}$qiaoxi.zhu@uts.edu.au, $^{4}$lucayongxu@tencent.com
}
\begin{document}

\maketitle

\begin{abstract}
  %Sound field decomposition is a task to predict waveforms in arbitrary directions by using a limited number of microphone signals as inputs. Sound field decomposition is fundamental to downstream tasks, including source localization, source separation, and sound field reproduction. Conventional sound field decomposition methods such as Ambisonics have limited spatial decomposition resolution. This paper proposes a learning-based Neural Sound field Decomposition (NeSD) framework to allow sound field decomposition with fine spatial direction resolution using recordings from microphone capsules of a few microphones at arbitrary positions. The inputs of a NeSD system include microphone signals, microphone positions, and queried directions. The outputs of a NeSD include the waveform and the presence probability of a queried position. We model the NeSD systems with different neural networks, including fully connected, time delay, and recurrent neural networks, respectively. We show that the NeSD systems outperform conventional Ambisonics and DOANet methods in sound field decomposition and source localization on speech, music, and sound events datasets. We release the source code at\footnote{\url{https://drive.google.com/file/d/14CnLZyYofIU-VvsHDNZZrS3cnCPaG9le/view} (for review)} and the demo at\footnote{\url{https://drive.google.com/file/d/1TfVQ8CrruBdP9CmZF-00ZNk7G7vlEfY_/view} (for review)}.
  
  %QZ:
  Sound field decomposition predicts waveforms in arbitrary directions using signals from a limited number of microphones as inputs. Sound field decomposition is fundamental to downstream tasks, including source localization, source separation, and spatial audio reproduction. Conventional sound field decomposition methods such as Ambisonics have limited spatial decomposition resolution. This paper proposes a learning-based Neural Sound field Decomposition (NeSD) framework to allow sound field decomposition with fine spatial direction resolution, using recordings from microphone capsules of a few microphones at arbitrary positions. The inputs of a NeSD system include microphone signals, microphone positions, and queried directions. The outputs of a NeSD include the waveform and the presence probability of a queried position. We model the NeSD systems respectively with different neural networks, including fully connected, time delay, and recurrent neural networks. We show that the NeSD systems outperform conventional Ambisonics and DOANet methods in sound field decomposition and source localization on speech, music, and sound events datasets. Demos are available at\footnote{\url{https://www.youtube.com/watch?v=0GIr6doj3BQ}}.
  %We release the source code at\footnote{\url{https://drive.google.com/file/d/14CnLZyYofIU-VvsHDNZZrS3cnCPaG9le/view} {\color{red}{(for review)}} }and the demo at\footnote{\url{https://drive.google.com/file/d/1TfVQ8CrruBdP9CmZF-00ZNk7G7vlEfY_/view} {\color{red}{(for review)}}}.

\end{abstract}

\section{Introduction}

%A sound field is a region in space in which sound waves are propagating. 
A sound field of a spatial region may contain sound waves propagating from different directions. Sound field decomposition \cite{rafaely2004plane, bernschutz2016microphone, koyama2018sparse, zhu2021experimental} %is a task to decompose waveforms recorded from microphone arrays to waveforms at a receiver in arbitrary directions. 
decomposes wave fields from arbitrary directions from signals recorded by microphone arrays.
This work introduces a learning-based neural sound decomposition (NeSD) approach to predict \textit{what}, \textit{where}, and \textit{when} %are the sources 
are sound in a recording. The NeSD system can be used as pre-processing for downstream tasks, %including 
such as sound localization and direction of arrival estimation \cite{adavanne2018sound, politis2020overview, cao2019two, grondin2019sound, qin2015generalized, xiao2015learning}, sound event detection \cite{mesaros2016tut, cakir2017convolutional, kong2020panns}, source separation \cite{stoter2019open, defossez2019music, seetharaman2019class, xu2013experimental, luo2019conv}, beamforming \cite{huang2019fast, alkhateeb2018deep, xu2020neural, zhang2021adl}, sound field reproduction \cite{betlehem2005theory, donley2018multizone, han2018two}, and augmented reality (AR) and virtual reality (VR) \cite{vorlander2015virtual, zotter2019ambisonics, liu2018relationship}.

A sound field consists of sounds coming from different directions. For example, musicians in a band may have different locations on a stage. Multiple speakers may have different locations in a room. Sound localization \cite{politis2020overview}, also called direction of arrival (DOA) estimation, is a task to predict the locations of sources. Previous DOA methods include parametric-based methods such as time difference of arrival (TDOA) \cite{huang2001real} and multiple signal classification (MUSIC) \cite{schmidt1986multiple}. Recently, neural networks have been introduced to address the DOA problem, such as convolutional recurrent neural networks (CRNNs) \cite{adavanne2018sound, grondin2019sound} and two stages estimation methods \cite{cao2019two}. However, many conventional DOA methods require the number of microphones to be larger than the number of sources and can not decompose waveforms.

% Qiaoxi 
% A sound field consists of sounds coming from different directions, such as musicians in a band at different locations on a stage or multiple speakers at different locations in a room. It posts challenges in sound recording and separation with a compact system. Sound localization \cite{politis2020overview} includes parametric-based methods such as time difference of arrival (TDOA) \cite{huang2001real} and multiple signal classification (MUSIC) \cite{schmidt1986multiple}. Recently, neural networks have been introduced to address the DOA problem, such as convolutional recurrent neural networks (CRNNs) \cite{adavanne2018sound, grondin2019sound} and two stages estimation methods \cite{cao2019two}. However, many conventional DOA methods require the number of microphones to be larger than the number of sources and can not decompose waveforms.

Another challenge of sound field decomposition is to separate signals in the sound field from different directions. Beamforming \cite{liu2010wideband, heymann2016neural, xu2020neural, zhang2021adl} is a technique to filter signals in specific beams. Conventional beamforming methods include the minimum variance distortionless response (MVDR) \cite{habets2009new}. Recently, neural network-based beamforming methods have been proposed for beamforming \cite{xu2020neural, zhang2021adl}. However, those beamforming methods do not predict the localization of sources. Previous neural network-based beamforming methods focused on on speech and were not train on the general sound sounds. Sound field decomposition is also related to the source separation problem, where deep neural networks have been applied to address the source separation problem \cite{luo2019conv, stoter2019open, wang2018supervised, huang2015joint, defossez2019music}. However, many source separation systems do not separate correlated waveforms from different directions. Recently, unsupervised source separation methods \cite{drude2019unsupervised, wisdom2020unsupervised, hu2012unsupervised} were proposed to separate unseen sources. Still, those methods do not separate highly correlated sources and do not predict the directions of sources. 

The conventional way of sound field decomposition requires numerous measurements to capture a sound field. First-order Ambisonics (FOA) \cite{gerzon1973periphony, furness1990ambisonics, frank2015producing} and high-order Ambisonics (HOA) \cite{daniel2004further, moreau20063d} were proposed to record sound fields. Ambisonics provides truncated spherical harmonic decomposition of a sound field. A $K$-th order ambisonics require $ (K+1)^{2} $ channels to record a sound field. A sound field is hard to record and process when $K$ is large. Moreover, the accurate reproduction in a head-sized volume up to 20 kHz would require an order $ K $ of more than 30. Recently, DOANet \cite{adavanne2018direction} was proposed to predict pseudo-spectrum, while does not predict waveforms. Still, here is a lack of works on neural network-based sound field decomposition that can achieve super-resolution of a sound field.

In this work, we propose a NeSD framework to address the sound field decomposition problem. The NeSD approach is inspired by the neural radiance fields (NeRF) \cite{mildenhall2020nerf} for view synthesis in computer vision. NeSD has the advantage of predicting the locations of wideband, non-stationary, moving, and arbitrary number of sources in a sound field. NeSD supports any layout of microphone array types, including uniform or non-uniform arrays such as planar, spherical array, or other irregular arrays. NeSD also supports the microphones to have different directivity patterns. NeSD can separate correlated signals in a sound field. NeSD can decompose a sound field with arbitrary spatial resolutions and can achieve better directivity than FOA and HOA methods. In training, the inputs to a NeSD system include the signals of arbitrary microphone arrays, the positions of all microphones, and arbitrary queried directions on a sphere. In inference, all sound field directions are input to the trained NeSD in mini-batches to predict the waveforms and the presence probabilities of sources in a sound field.

This work is organized as follows. Section \ref{section:problem_definition} introduces the sound field decomposition problem. Section \ref{section:nesd} introduces our proposed NeSD framework. Section \ref{section:experiments} shows experiments. Section \ref{section:conclusion} concludes this work.

\section{Problem Statement}\label{section:problem_definition}

%We first define the sound field decomposition problem. 
The signals recorded from a microphone capsule are denoted as $ \textbf{x} = \{x_{1}(t), ..., x_{M}(t)\} $, where $ M $ is the number of microphones in the capsule. The $m$-th microphone signal is $ x_{m}(t) \in \mathbb{R}^{T} $ where $ T $ is the number of samples in an audio segment. The microphone capsule can be any type, such as the Ambisonics or circular capsule.

The coordinate of the $m$-th microphone in the spherical coordinate system is $q_{m}(t)$, where $ q_{m}(t) = \{r_{m}(t), \theta_{m}(t), \phi_{m}(t)\} $ denotes the distance, azimuthal angle, and polar angle of the microphone. The position information of all microphones is $ \mathbf{q} = \{q_{1}(t), ..., q_{M}(t)\} $. Note that $ q_{m}(t) $ is a time-dependent variable to reflect moving microphones. For a static microphone, all of $r_{m}(t)$, $\theta_{m}(t)$,  and $\phi_{m}(t)$ in $ q_{m}(t) $ have constant values.

We denote a continuous sound field as $ \mathbf{s} = s(\Omega, t) \in \mathbb{S}^{2} \times \mathbb{R}^{T} $ where $ \mathbb{S}^{2} $ is a sphere. Each direction $ \Omega $ is described by an azimuth angle $ \theta \in [0, 2 \pi) $ and a polar angle $ \phi \in [0, \pi] $. Sound field decomposition is a task to estimate $ \mathbf{s} $ from the microphone signals $ \mathbf{x} $ and the microphone positions $ \mathbf{q} $:
\begin{equation} \label{eq:mapping_org}
\hat{s}(\Omega, t) = f_{\mathbb{S}^{2}}(\mathbf{x}, \mathbf{q}),
\end{equation}
\noindent where $ f_{\mathbb{S}^{2}}(\cdot, \cdot) $ is a sound field decomposition mapping and $ \hat{s}(\Omega, t) $ is the estimated %sound field.
waveform.

\section{Neural Sound Field Decomposition (NeSD)}\label{section:nesd}

We introduce the training, the data creation, the NeSD architecture, the hard example mining, and the inference of NeSD in this section.

\subsection{Empirical Risk}
Directly model $ f_{\mathbb{S}^{2}} $ is intractable because the output dimension $ \mathbb{S}^{2} \times \mathbb{R}^{L} $ is infinite. Instead, we propose a mapping $ f $ to predict $ \hat{s}(\Omega, t) $ conditioned on a direction $ \Omega $ for all $ \Omega \in \mathbb{S}^{2} $:
\begin{equation} \label{eq:mapping}
\hat{s}(\Omega, t) = f(\mathbf{x}, \mathbf{q}, \Omega),
\end{equation}
We model $ f $ in (\ref{eq:mapping}) by using a neural network. We denote the risk $ l $ between the estimated sound field $ \mathbf{\hat{s}} $ and the oracle sound field $ \mathbf{s} $ as:
\begin{equation} \label{eq:risk}
l = \mathbb{E}_{\mathbf{s} \sim p_{\text{field}}(\mathbf{s})}\mathbb{E}_{\Omega \sim p_{\text{sphere}}(\Omega)} \mathbb{E}_{t \sim p_{\text{time}}(t)} d(s(\Omega, t), \hat{s}(\Omega, t)),
\end{equation}
\noindent where $ p_{\text{field}} $, $ p_{\text{sphere}} $, and $ p_{\text{time}} $ are the distributions of sound field $ \mathbf{s} $, direction $ \Omega $, and time $ t $, respectively. The loss function is denoted by $ d(\cdot, \cdot) $. Equation (\ref{eq:risk}) shows that the risk $ l $ consists of expectations over $ p_{\text{field}} $, $ p_{\text{sphere}} $, and $ p_{\text{time}} $. However, directly optimize the risk (\ref{eq:risk}) is intractable. To address this problem, we propose to minimize the empirical risk in mini-batches:
\begin{equation} \label{eq:emp_risk}
l_{\text{batch}} = \sum_{\substack{\mathbf{s} \sim p_{\text{field}}(\mathbf{s}) \\ b=1}
}^{B} 
\sum_{\substack{\Omega \sim p_{\text{sphere}}(\Omega) \\ q=1}}^{Q} \sum_{t=1}^{T} d(s(\Omega, t), \hat{\Omega, s}(t)),
\end{equation}
\noindent where $ B $ is the mini-batches number to sample sound fields from a dataset and $Q$ is the number of directions to sample on a sphere. In each mini-batch, sound field signals $ \mathbf{s} $ are sampled from $ p_{\text{field}}(\mathbf{s}) $. Directions $ \Omega $ are sampled from $ p_{\text{sphere}}(\Omega) $. By this means, the optimization of (\ref{eq:emp_risk}) is tractable. The risk (\ref{eq:emp_risk}) is differentiable with respect to the learnable parameters of $ f $, so that the learnable parameters can be optimized by gradient-based methods. 

\subsection{Create Microphone Sound Field Signals}

The optimization of (\ref{eq:emp_risk}) requires paired microphone signals $ \mathbf{x} $, microphone positions information $ \mathbf{q} $ and sound field signals $ \mathbf{s} $ for training. However, it is impossible to obtain oracle sound field signals $ \mathbf{s} $ in real scenes. To address this problem, we propose to create microphone and sound field signals from point sources.

We create $ I $ far field point sources $ \{ a_{i}(t) \}_{i=1}^{I} $ from $ I $ randomly sampled directions $ \{ \Omega_{i} \}_{i=1}^{I} $. Each $ a_{i}(t) $ is a randomly selected monophonic audio segment from an audio dataset such as a speech or a music dataset. In a free field without room reverberation, the sound field $ \mathbf{s} $ can be created by:
\begin{wrapfigure}{r}{0.5\linewidth}
\centerline{\includegraphics[width=0.5\columnwidth]{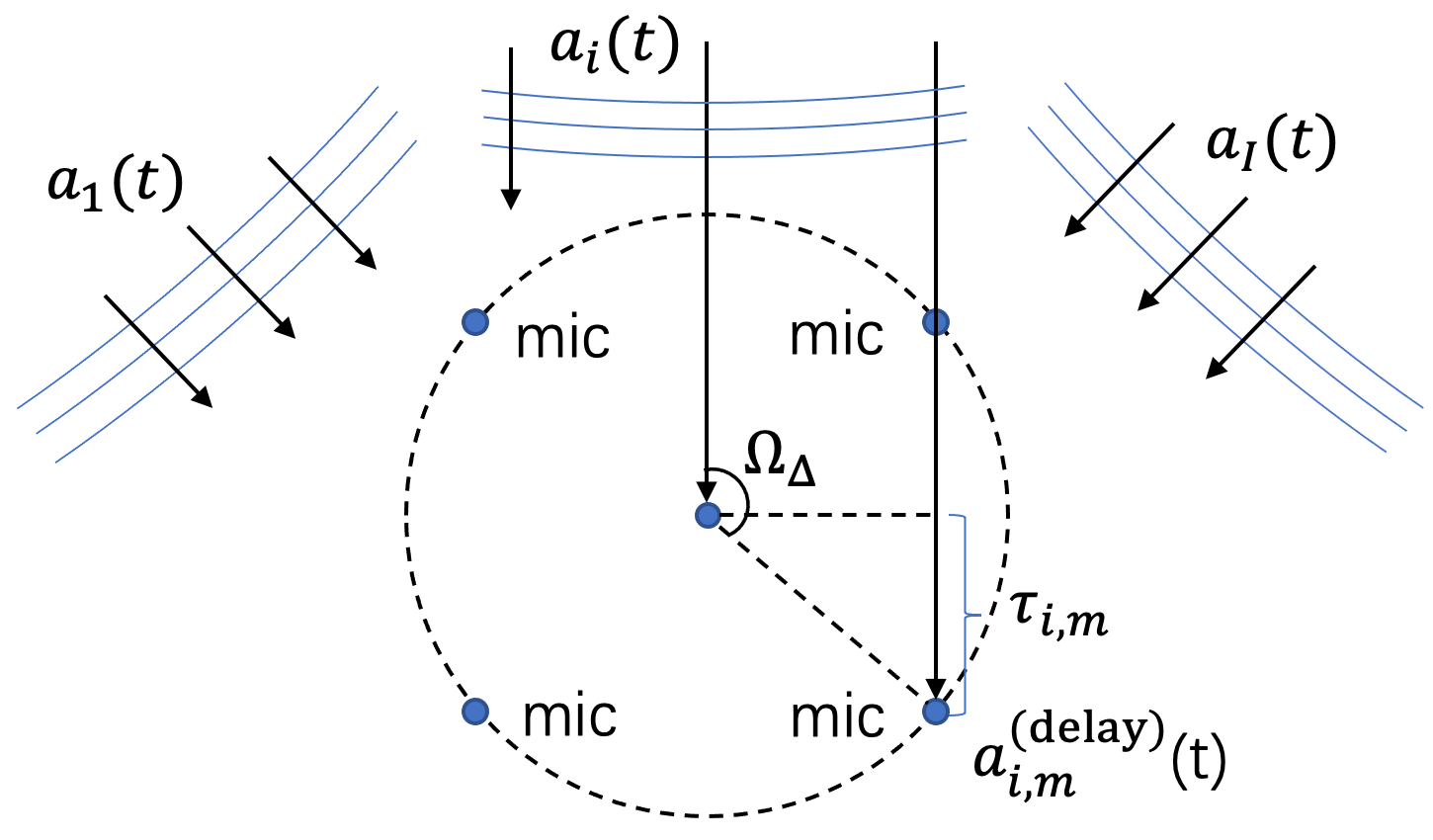}}
%\rule{0.5\linewidth}{0.5\linewidth}
\caption{Point source signals arrive microphones. \\}
\label{fig:mic}
\end{wrapfigure}
\begin{equation} \label{eq:construct_sound_field}
s(\Omega, t) = \left\{\begin{matrix}
a_{i}(t), &\Omega=\Omega_{i} & \\ 
0, &\Omega = \mathbb{S}^{2} / \{\Omega_{i}\}_{i=1}^{I}. &
\end{matrix}\right.
\end{equation}
In (\ref{eq:construct_sound_field}), a sound field $ \mathbf{s} $ only has non-zero values in the directions containing point sources.

Next, we create microphone signals $ \mathbf{x} $. First, all $ I $ point source signals are propagated to all $ M $ microphones. We denote $ a_{i, m}^{\text{(delay)}}(t) $ as the $ i $-th point source signal arrives the $ m $-th microphone:
\begin{equation} \label{eq:construct_mic_delay}
a_{i, m}^{\text{(delay)}}(t) = a_{i}(t - \tau_{i, m}),
\end{equation}
\noindent where $ \tau_{i, m} $ is the propagation time as shown in Fig. \ref{fig:mic}. The propagation time $ \tau_{i, m} $ can be calculated by the speed of sound $ c $, the distance from the $m$-th microphone to the origin $ r_{m} $, and the included angle $ \Omega_{\Delta} $ between the $ i $-th point source and the $ m $-th microphone by: $ \tau_{i, m} = - r_{m} \cos \Omega_{\Delta} $ as shown in Fig. \ref{fig:mic}.

Microphones have different directivity patterns. So the signal recorded by a microphone can be different from the signal arrives the microphone. We denote the $i$-th signal recorded by the $m$-th microphone as:
\begin{equation} \label{eq:construct_mic_response}
a_{i, m}^{\text{(mic)}}(t) = g(a_{i, m}^{\text{(delay)}}(t), \Omega_{\Delta}),
\end{equation}
\noindent where $ g $ is a transfer function. To simplify $g$, we assume the directivity pattern is constant for all frequencies. Then, equation (\ref{eq:construct_mic_response}) equals $ a_{i, m}^{\text{(delay)}}(t) $ for an omni microphone and equals $ a_{i, m}^{\text{(delay)}}(t) (1 + \cos \Omega_{\Delta}) / 2 $ for a cardioid microphone. Then, the microphone signals of all point sources $ a_{i, m}^{(\text{mic})}(t) $ are summed to constitute the $m$-th microphone signal:
\begin{equation} \label{eq:construct_mic_sum}
x_{m}(t) = \sum_{i=1}^{I}a_{i, m}^{\text{(mic)}}(t)
\end{equation}
The microphone positions information $ \textbf{q} $ can be obtained from the pre-defined microphone array. By this means, we obtain paired $ \mathbf{x} $, $ \textbf{q} $, and $ \mathbf{s} $ to train the NeSD system.

\subsection{NeSD Input Embeddings}
\begin{figure}[t]
  \centering
  \centerline{\includegraphics[width=\columnwidth]{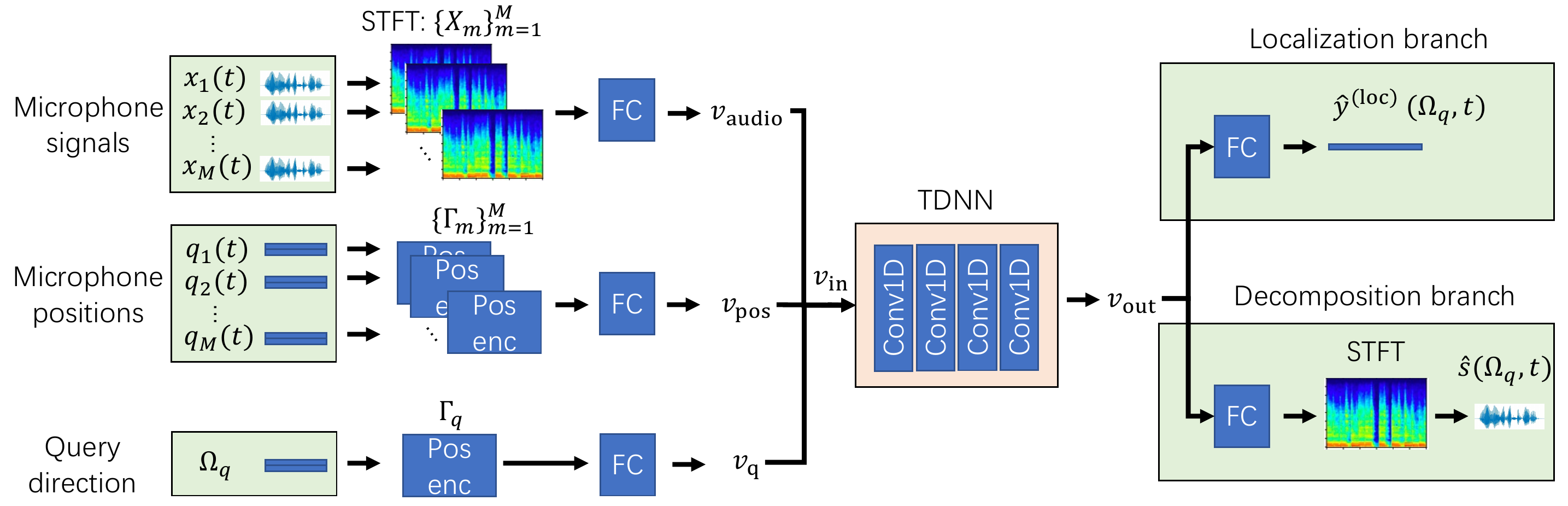}}
  \caption{Framework of the NeSD system.}
  \label{fig:framework}
\end{figure}

The core part of a NeSD system is to build the mapping in (\ref{eq:mapping}). The input to a NeSD system includes the microphone signals $ \mathbf{x} $, the microphone positions information $ \textbf{q} $, and queried directions $ \Omega_{\text{q}} $. Fig. \ref{fig:framework} shows the NeSD framework. We describe the submodules of a NeSD system as follows.

\subsubsection{Audio Embedding}\label{section:mic_embedding}
 We apply short-time Fourier transform (STFT) on each microphone signal $ x_{m}(t) $ to extract the STFT of waveforms. Each STFT $ X_{m} $ has a shape of $ L \times F $, where $ L $ and $ F $ are frames number and frequency bins number, respectively. We concatenate $M$ STFTs $ \{ X_{m} \}_{m=1}^{M} $ along the frequency axis to constitute $ X_{\text{concat}} $ with a shape of $ L \times FM $. Then, we apply a fully connected layer on $ X_{\text{concat}} $ to calculate an \textit{audio embedding} $ v_{\text{audio}} $ with a shape of $ L \times C $:
\begin{equation} \label{eq:sp_emb}
v_{\text{audio}} = X_{\text{concat}}W_{\text{audio}} + b_{\text{audio}},
\end{equation}
\noindent where $ C $ is the hidden units number. $ W_{\text{audio}} $ and $ b_{\text{audio}} $ are learnable weights and biases with shapes of $ MF \times C $ and $ C $, respectively. The audio embedding $ v_{\text{audio}} $ contains the content information of microphone signals.

\subsubsection{Microphone Positions Embedding}
We map the microphone positions to a \textit{microphone positions embedding} $ v_{\text{pos}} $ which contains the position information of all microphones. We find that directly construct $ v_{\text{pos}} $ from $ \theta_{m}(t) $ and $ \phi_{m}(t) $ leads to poor performance in sound field decomposition. To address this problem, we propose to apply positional encoding \cite{mildenhall2020nerf} on $ \theta_{m}(t) $ and $ \phi_{m}(t) $:
\begin{equation} \label{eq:pos_enc}
\begin{split}
& \gamma(\theta) = \{ \sin(2^0 \pi \theta), \cos(2^0 \pi \theta), ..., \sin(2^{P - 1} \pi \theta), \cos(2^{P - 1} \pi \theta) \} \\
& \gamma(\phi) = \{ \sin(2^0 \pi \phi), \cos(2^0 \pi \phi), ..., \sin(2^{P - 1} \pi \phi), \cos(2^{P - 1} \pi \phi) \},
\end{split}
\end{equation}
\noindent where the subscript $m$ and $t$ are omitted for (\ref{eq:pos_enc}) for conciseness and $ P $ is a hyper-parameter. Larger $ P $ indicates better spatial resolution. Positional encodings have the advantage of remaining high frequency of spatial angles and is beneficial for high-resolution sound field decomposition. We denote the position encoding of the $m$-th microphone as $ \Gamma_{m} = \{ \gamma(\theta), \gamma(\phi) \} $ with a shape of $ L \times 4P $, where the number 4 includes the cosine and sine of the azimuth angle and polar angle. We concatenate all $ M $ position encodings along the feature dimension to obtain $ \Gamma_{\text{concat}} $ which has a shape of $ L \times 4PM $. Then, we apply a fully connected layer on $ \Gamma_{\text{concat}} $ to calculate a microphone positions embedding $ v_{\text{pos}} $ with a shape of $ L \times C $:
\begin{equation} \label{eq:topo_emb}
v_{\text{pos}} = \Gamma_{\text{concat}} W_{\text{pos}} + b_{\text{pos}},
\end{equation}
\noindent where $ W_{\text{pos}} $ and $ b_{\text{pos}} $ are learnable weights and bias with shapes of $ 4MP \times C $ and $ C $, respectively.

\subsubsection{Query Direction Embedding}

A query direction $ \Omega_{\text{q}} $ is used to condition a NeSD in (\ref{eq:mapping}) to predict the decomposed waveform $ \hat{s}(\Omega_{\text{q}}, t) $ in direction $ \Omega_{\text{q}} $. Similar to the microphone position encoding in (\ref{eq:pos_enc}), we map $ \Omega_{q} $ to a position embedding $ \Gamma_{q} $ with a shape of $ L \times 4P $. Then, we apply a fully connected layer on $ \Gamma_{q} $ to calculate a \textit{query direction embedding} $ v_{\text{q}} $ with a shape of $ L \times C $:
\begin{equation} \label{eq:query_emb}
v_{\text{q}} = \Gamma_{\text{q}} W_{\text{q}} + b_{\text{q}},
\end{equation}
\noindent where $ W_{\text{p}} $ and $ b_{\text{p}} $ are learnable weights and bias with shapes of $ 4P \times C $ and $ C $, respectively. The query direction embedding $ v_{\text{q}} $ contains high spatial resolution information of the query direction $ \Omega_{\text{q}} $.

Then, we concatenate all the audio embedding $ v_{\text{audio}} $, the microphone positions embedding $ v_{\text{pos}} $, and the query direction embedding $ v_{\text{q}} $ along the feature dimension to calculate an input embedding $ v_{\text{in}} $ with a shape of $ T \times 3C $. The input embedding $ v_{\text{in}} $ contains all information of microphone signals, the microphone positions, and query directions.

\subsection{NeSD Backbone Networks}
We apply a time delay neural network (TDNN) \cite{waibel1989phoneme} on $ v_{\text{in}} $ to extract the high-level information of microphone signals, microphone positions, and query directions. The TDNN consists of four time-domain one-dimensional convolutional layers. Each convolutional layer outputs $ H $ channels followed by a batch normalisation \cite{ioffe2015batch} and a ReLU non-linearity \cite{brownlee2019gentle}. A TDNN has the advantage of capturing time-dependent information. In a special case when the kernel sizes of convolutional layers are set to 1, the convolutional layers are reduced to fully connected layers. 
We denote the output of the backbone network is as $ v_{\text{out}} $ with a shape of $ T \times H $. In addition, we implement a 3-layer fully connected neural network and a 3-layer bi-directional gated recurrent unit (GRU) \cite{chung2014empirical} as backbone networks for comparison.

\subsection{NeSD Output Branches}
We build multiple output branches on the backbone network output $ v_{\text{out}} $ to support multiple tasks, including sound localization and decomposition.

\subsubsection{Sound Localization Branch}
In the sound localization branch shown in Fig. \ref{fig:framework}, we apply a fully connected layer on $ v_{\text{out}} $ followed by a sigmoid nonlinearity to predict the presence probability of sources in direction $ \Omega_{\text{q}} $:
\begin{equation} \label{eq:loc_fc}
\hat{y}_{\text{loc}}(\Omega_{\text{q}}, n) = \sigma(v_{\text{out}} V_{\text{loc}} + b_{\text{loc}}),
\end{equation}
\noindent where $ n $ is the frame index, and $ V_{\text{loc}} $ and $ b_{\text{loc}} $ are learnable weights and biases with shapes of $ C \times 1 $ and $ 1 $, respectively. The output probability $ \hat{y}_{\text{loc}}(\Omega_{\text{q}}, n) $ indicates the probability of point sources in direction $ \Omega_{\text{q}} $.

\subsubsection{Source Decomposition Branch}
In the source decomposition branch shown in Fig. \ref{fig:framework}, we apply a fully connected layer on $ v_{\text{out}} $ followed by a sigmoid nonlinearity to predict the ideal ratio mask \cite{narayanan2013ideal} of the input omni waveform in the time-frequency domain, where the omni waveform is calculated by averaging all microphone signals. The time-frequency mask in direction $ \Omega_{\text{q}} $ is predicted by:
\begin{equation} \label{eq:ss_fc}
M(\Omega_{\text{q}}, n, f) = \sigma(v_{\text{out}} V_{\text{ss}} + b_{\text{ss}}),
\end{equation}
\noindent where $n$ and $f$ are frame index and frequency bin index, respectively. Matrices $ V_{\text{ss}} $ and $ b_{\text{ss}} $ are learnable weights and biases with shapes of $ C \times 1 $ and $ 1 $, respectively. The time-frequency mask $ M(\Omega_{\text{q}}, n, f) $ has a shape of $ T \times F $ and is multiplied with the STFT of the omni waveform $ X_{\text{omni}}(n, f) $ followed by an inverse STFT to obtain the separated waveform $ \hat{s}(\Omega_{\text{q}}, t) $ at $ \Omega_{\text{q}} $:
\begin{equation} \label{eq:ss_istft}
\hat{s}(\Omega_{\text{q}}, t) = \mathcal{F}^{-1}(M(\Omega_{\text{q}}, n, f) \odot X_{\text{omni}}(n, f)),
\end{equation}
\noindent where $ \mathcal{F}^{-1} $ is the inverse STFT, and $ \odot $ is the element-wise multiplication.

\subsection{Loss Function}
The loss function $ d $ in (\ref{eq:emp_risk}) consists of a localization loss $ l_{\text{loc}} $ and a decomposition loss $ l_{\text{ss}} $:
\begin{equation} \label{eq:loss_full}
d = l_{\text{loc}} + l_{\text{ss}},
\end{equation}
\noindent where $ l_{\text{loc}} $ is the binary cross-entropy between the predicted presence probability $ \hat{y}_{\text{loc}}(\Omega_{\text{q}}, n) $ and the ground truth $ y_{\text{loc}}(\Omega_{\text{q}}, n) $ in a queried direction $ \Omega_{\text{q}} $:
\begin{equation} \label{eq:loss_loc}
l_{\text{loc}} = l_{\text{bce}}(\hat{y}_{\text{loc}}(\Omega_{\text{q}}, n), y_{\text{loc}}(\Omega_{\text{q}}, n))
\end{equation}
The source decomposition loss is the mean absolute error between the predicted waveform $\hat{s}(\Omega_{\text{q}}, t)$ and the ground truth waveform $s(\Omega_{\text{q}}, t)$ in a queried direction $ \Omega_{\text{q}} $:
\begin{equation} \label{eq:loss_ss}
l_{\text{ss}} = || \hat{s}(\Omega_{\text{q}}, t) - s(\Omega_{\text{q}}, t) ||_{1},
\end{equation}

\subsection{Hard Example Mining}

\begin{wrapfigure}{r}{0.5\linewidth}
\centerline{\includegraphics[width=0.5\columnwidth]{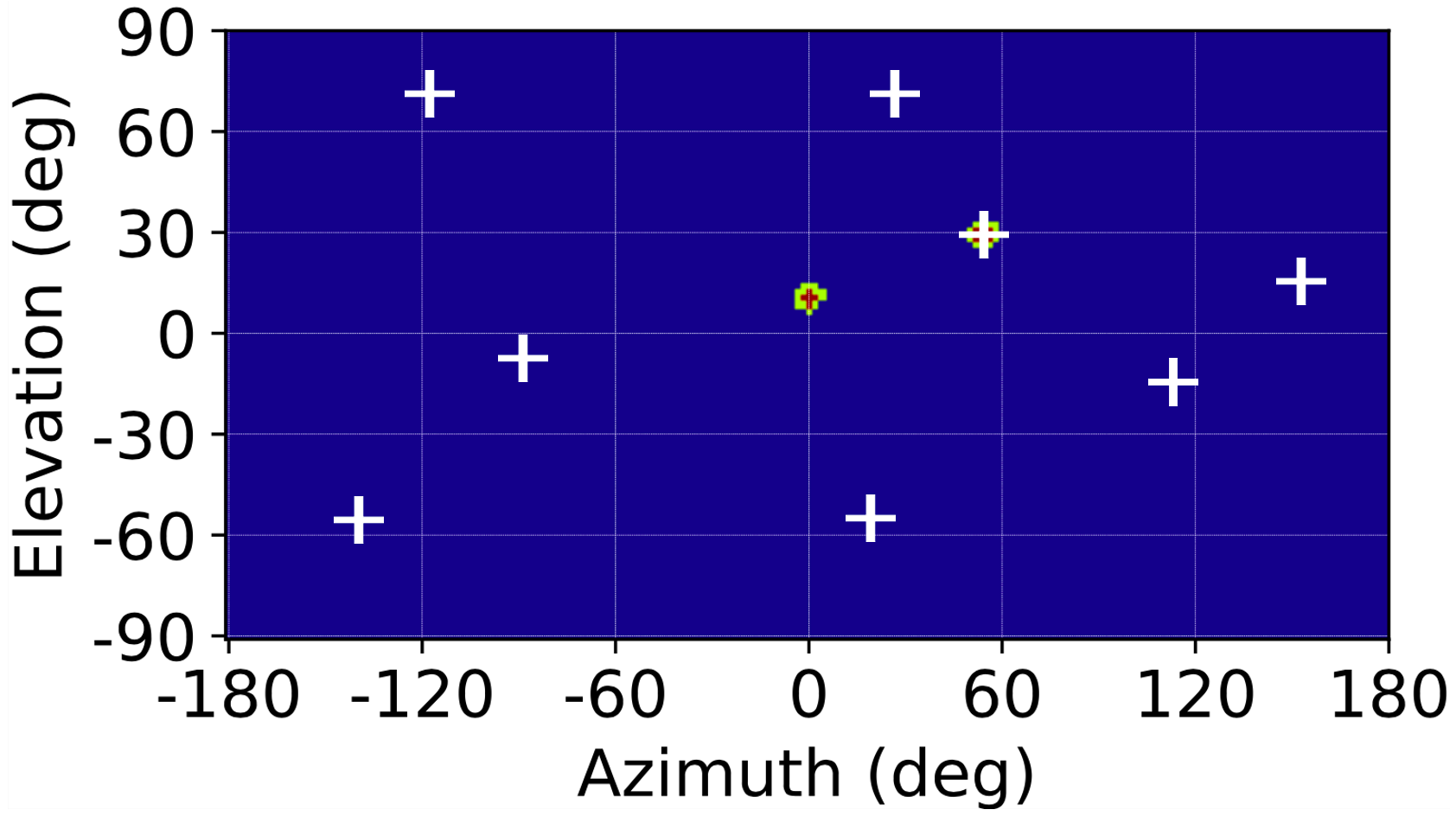}}
\caption{The cross symbols show the sampled query directions by using Algorithm \ref{alg:hard_example_mining}.}
\label{fig:hard_example_mining}
\end{wrapfigure}
The empirical loss (\ref{eq:emp_risk}) requires the sampling of $ \mathbf{s} $ from $ p_{\text{field}}(\mathbf{s}) $ and the sampling of query directions $ \Omega_{\text{q}} $ from $ p_{\text{sphere}}(\Omega) $. On one hand, we uniformly sample audio segments $ a_{i}(t) $ from an audio dataset and process $ a_{i}(t) $ to $ \mathbf{s} $ by (\ref{eq:construct_sound_field}). On the other hand, we need to elaborately sample query directions $ \Omega_{q} $ to avoid the situation where all query directions are silent. We propose a hard mining strategy in Algorithm \ref{alg:hard_example_mining} to increase the proportion of non-silent query directions among $ Q $ sampled directions. Fig. \ref{fig:hard_example_mining} shows an example of $Q$ sampled query directions by Algorithm \ref{alg:hard_example_mining}. The cross symbols are sampled query directions. The red dots are the positions of point sources.

\begin{algorithm}[t]
  \caption{Hard example mining of one query direction.}\label{alg:hard_example_mining}
  \begin{algorithmic}
    \State \textbf{Inputs}: $ \mathbb{S}^{2} $, point source directions $ \{ \Omega_{i} \}_{i=1}^{I} $, $ p_{\text{thres}}=0.9. $
    \State \textbf{Outputs}: a query direction $ \Omega_{\text{q}} $.
    \State Sample $ p \sim \text{Uniform}(0, 1) $
    \If{$ p < p_{\text{thres}} $}
      \State Sample $ \Omega_{\text{q}} \sim \mathbb{S}^{2} $.
    \Else
      \State Sample $ \Omega_{\text{q}} \sim \{ \Omega \}_{i=1}^{I} $.
    \EndIf
  \end{algorithmic}
\end{algorithm}

\begin{algorithm}[t]
  \caption{Training of a NeSD system.}\label{alg:training}
  \begin{algorithmic}[1]
    \State \textbf{Inputs}: An audio dataset $ \mathbb{D} $. Microphone positions $ \mathbf{q} $. $B$: mini-batch size, $I$: sources number, $Q$: query directions number.
    \State \textbf{Outputs}: Optimized parameters of a NeSD system.
    \While{not converge}
    \For{$b = 1, ..., B$} \
      \State Sample $ \{a_{i}(t)\}_{i=1}^{I} $ uniformly from $\mathbb{D}$.
      \State Constitute $ \mathbf{s} $ by using (\ref{eq:construct_sound_field}) and $ \mathbf{x} $ by using (\ref{eq:construct_mic_sum}).
      \For{$ q = 1, ..., Q $} \
        \State Sample a query direction $\Omega_{q} $ by using Algorithm \ref{alg:hard_example_mining}.
      \EndFor
      \State Constitute the $b$-th training data: $ \{ \mathbf{x}, \mathbf{q}, \{ \Omega_{\text{q}} \}_{q=1}^{Q} \} $
    \EndFor
    \State Calculate empirical loss by (\ref{eq:emp_risk}) and (\ref{eq:loss_full}). Calculate the gradient with respect to the learnable parameters of NeSD. 
    \State Update the learnable parameters with gradient-based optimization methods.
    \EndWhile
  \end{algorithmic}
\end{algorithm}

\subsection{Inference}
Algorithm \ref{alg:training} summarizes the training of a NeSD system. At inference, we input the microphone signals $ \mathbf{x} $, the microphone positions information $ \mathbf{q} $, and any query directions $ \Omega_{\text{q}} $ to the trained NeSD system. For the sound field decomposition purpose, we input all azimuth angle and polar angle combinations to the NeSD system. The azimuth angles and polar angles have a granularity of $\kappa$ degrees. The granularity $\kappa$ can be tuned in the inference stage. Lower $\kappa$ leads to higher spatial decomposition resolution while requires higher computation cost. Algorithm \ref{alg:inference} summarizes the inference of a NeSD system.

\begin{algorithm}[t]
  \caption{Sound field decomposition with a trained NeSD system.}\label{alg:inference}
  \begin{algorithmic}[1]
    \State \textbf{Inputs}: $ \mathbf{x} $, $ \mathbf{q} $. 
    \State \textbf{Outputs}: $\mathbf{s}$
    \For{$ \Omega \in \mathbb{S}^{2} $}
      \State $ \hat{s}(\Omega, t) = f(\mathbf{x}, \mathbf{q}, \Omega)$ 
    \EndFor
  \end{algorithmic}
\end{algorithm}

\section{Experiments}\label{section:experiments}

\subsection{Datasets}

We experiment the NeSD system across a variety of datasets including speech, music, and sound events datasets. We apply the VCTK dataset \cite{yamagishi2019cstr} containing 108 speakers as the speech dataset. The VCTK dataset is split into 100 speakers for training and 8 speakers for testing. We apply the MUSDB18 dataset \cite{rafii2017musdb18} containing 150 songs as the music dataset. The MUSDB18 dataset is split into 100 songs for training and 50 songs for testing. Each song contains four individual tracks including vocals, bass, drums, and other. We apply the NIGENS dataset \cite{trowitzsch2019nigens} containing 14 distinguished sound classes as the sound events dataset. 

We resample all audio recordings into 16 kHz and remove all silent frames. Then, audio recordings are split into 3-second segments. We apply a STFT with a window size of 32 ms and a hop size of 10 ms to calculate audio embeddings as described in Section \ref{section:mic_embedding}. We set the position encoding hyper-parameter $ P $ described in (\ref{eq:pos_enc}) to 5. The hidden units $ C $ is set to 128 and the number of output feature maps number $ H $ is set to 256. An Adam optimizer \cite{kingma2014adam} with a learning rate of 0.001 is used to train the NeSD system. The training takes 6 hours on a single Tesla V100 GPU card.

\subsection{Sound Field Energy}

\begin{figure}[t]
  \centering
  \centerline{\includegraphics[width=\columnwidth]{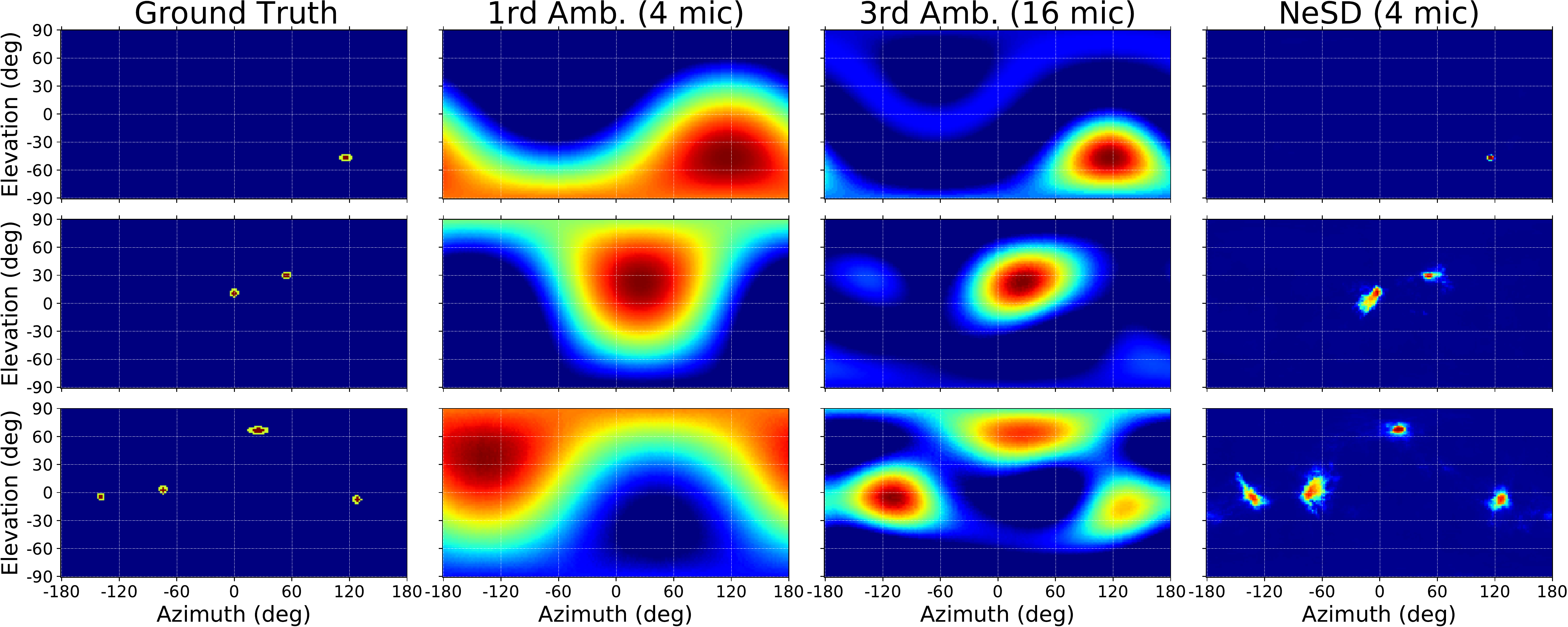}}
  \caption{Sound field decomposition. From left to right: ground truth sound field containing one to four speakers, sound field decomposition with 1-order Ambisonics, with 3-order Ambisonics, and with NeSD.}
  \label{fig:result_amb_nesd}
\end{figure}

First, we apply the mean absolute error between the predicted sound field energy and the ground truth sound field energy to evaluate the sound field decomposition performance:
\begin{equation} \label{eq:mae_metric}
MAE = \frac{1}{\sum_{\Omega \in \mathbb{S}^{2}}N}\sum_{\Omega \in \mathbb{S}^{2}}\sum_{n=1}^{N} | \hat{y}_{\text{loc}}(\Omega, n) - y_{\text{loc}}(\Omega, n) |
\end{equation}
Table \ref{tab:result_mae} shows the comparison of using Ambisonics methods and the NeSD methods to decompose sound fields. We experiment on sound fields containing 1, 2, 4, and 8 sources. The source types vary from speech, music, to sound events. Lower MAE indicates better performance. Table \ref{tab:result_mae} shows that all our proposed NeSD-DNN, NeSD-TDNN, and NeSD-GRU systems achieve lower MAE than Ambisonics-based methods. Specifically, the NeSD-GRU system achieves lower MAE than the NeSD-DNN and the NeSD-TDNN system, indicating that long-time dependency is beneficial for audio localization. The NeSD systmes with 4 microphones even surpass the 3-order Ambisonics system recorded with 16 microphones. The first column in Fig. \ref{fig:result_amb_nesd} shows the ground truth sound field. The second and third column show the 1-order and 3-order decomposition results. The fourth row shows the sound field predicted by NeSD. Fig. \ref{fig:result_amb_nesd} shows that the NeSD system achieves a better estimation of the ground truth sound field than Ambisonics decoding. The NeSD-GRU system achieves an MAE of 0.011 to 0.114 from 1 source to 8 sources evaluated on speech, compared with the 3-order Ambisonic method from 0.188 to 0.778.

\begin{table}
  \caption{Mean absolute error of sound field field decomposition.}
  \vspace{6pt}
  \label{tab:result_mae}
  \centering
  \resizebox{\textwidth}{!}{%
  \begin{tabular}{lcccccccccccccc}
    \toprule
    & & \multicolumn{4}{c}{\textbf{\textsc{Speech}}} & \multicolumn{4}{c}{\textbf{\textsc{Music}}} & \multicolumn{4}{c}{\textbf{\textsc{Sound event}}} \\
	\cmidrule(lr){3-6} \cmidrule(lr){7-10} \cmidrule(lr){11-14}
    & mic. & 1 src & 2 src & 4 src & 8 src & 1 src & 2 src & 4 src & 8 src & 1 src & 2 src & 4 src & 8 src \\
    \midrule
    Ambisonics 1 & 4 & 0.454 & 0.673 & 0.984 & 1.775 & 0.428 & 0.649 & 1.117 & 1.633 & 0.489 & 0.493 & 0.962 & 2.000 \\
    Ambisonics 3 & 16 & 0.188 & 0.271 & 0.401 & 0.778 & 0.168 & 0.267 & 0.470 & 0.679 & 0.214 & 0.300 & 0.451 & 0.768 \\
    \midrule
    NeSD-DNN & 4 & 0.095 & 0.240 & 0.314 & 0.372 & 0.075 & 0.258 & 0.295 & 0.328 & 0.094 & 0.278 & 0.294 & 0.313 \\ 
    NeSD-TDNN & 4 & 0.020 & 0.045 & 0.103 & 0.233 & 0.016 & 0.070 & 0.172 & 0.250 & 0.021 & 0.113 & 0.217 & 0.266 \\
    NeSD-GRU & 4 & \textbf{0.011} & \textbf{0.017} & \textbf{0.046} & \textbf{0.114} & \textbf{0.012} & \textbf{0.042} & \textbf{0.117} & \textbf{0.154} & \textbf{0.013} & \textbf{0.067} & \textbf{0.155} & \textbf{0.214}\\
    \bottomrule
\end{tabular}}
\end{table}

\subsection{Sound Field Decomposition}

We apply the signal-to-distortion for decomposition (SDRD) metric \cite{koyama2018sparse} to evaluate sound field decomposition. Different from conventional source decomposition metrics, SDRD evaluate the SDR in the entire sound field.
\begin{equation} \label{eq:sdrd_metric}
SDRD = 10 \log_{10}\frac{\sum_{\Omega \in \mathbb{S}^{2}}\sum_{t=1}^{T} | s(\Omega, t) |^{2}} {\sum_{\Omega \in \mathbb{S}^{2}}\sum_{t=1}^{T}| s(\Omega, t) - \hat{s}(\Omega, t) |^{2}}
\end{equation}
\begin{table}
  \caption{SDRD results (dB)}
  \vspace{6pt}
  \label{tab:result_sdrd}
  \centering
  \resizebox{\textwidth}{!}{%
  \begin{tabular}{lcccccccccccccc}
    \toprule
    & & \multicolumn{4}{c}{\textbf{\textsc{Speech}}} & \multicolumn{4}{c}{\textbf{\textsc{Music}}} & \multicolumn{4}{c}{\textbf{\textsc{Sound event}}} \\
	\cmidrule(lr){3-6} \cmidrule(lr){7-10} \cmidrule(lr){11-14}
    & mic. & 1 src & 2 src & 4 src & 8 src & 1 src & 2 src & 4 src & 8 src & 1 src & 2 src & 4 src & 8 src \\
    \midrule
    Ambisonics 0 & 1 & -33.68 & -32.41 & -33.55 & -32.12 & -33.64 & -33.11 & -29.37 & -31.59 & -30.17 & -30.88 & -32.04 & -32.93 \\
    Ambisonics 1 & 4 & -27.78 & -27.20 & -27.43 & -26.70 & -27.20 & -27.57 & -24.34 & -26.27 & -25.49 & -26.09 & -26.88 & -27.09 \\
    Ambisonics 3 & 16 & -22.00 & -21.62 & -21.15 & -21.31 & -21.25 & -21.66 & -19.43 & -20.81 & -20.73 & -20.95 & -21.49 & -21.46 \\
    \midrule
    NeSD-DNN & 4 & -2.86 & -10.55 & -11.91 & -9.34 & -5.45 & -12.78 & -13.00 & -10.32 & -2.91 & -11.21 & -12.29 & -11.38 \\
    NeSD-TDNN & 4 & -1.66 & \textbf{-4.84} & -7.14 & -6.77 & -3.96 & \textbf{-11.29} & \textbf{-11.87} & \textbf{-9.45} & \textbf{-1.32} & \textbf{-9.14} & \textbf{-12.13} & \textbf{-10.81} \\
    
    NeSD-GRU & 4 & \textbf{-0.80} & -6.11 & \textbf{-6.78} & \textbf{-4.67} & \textbf{-3.82} & -14.39 & -13.74 & -9.67 & -3.27 & -14.16 & -13.00 & -11.44 \\
    \bottomrule
\end{tabular}}
\end{table}

Table \ref{tab:result_sdrd} shows the comparison of SDRD using Ambisonics methods and the NeSD methods to decompose sound fields. The higher SDRD indicates the better performance. The 0-order Ambisonics is regarded as a baseline system which indicates no decomposition. The 0-order Ambisonics achieves SDRD of around -30 dB. Table \ref{tab:result_sdrd} shows that the NeSD systems significantly improve the Ambisonics methods by 20 to 30 dB. The NeSD-TDNN system and the NeSD-GRU system achieve similar results, both of which outperform the NeSD-DNN system, indicating that temporal information is important for sound field decomposition.

\subsection{Source Localization}
We evaluate the average angular error $E_{\text{DOA}}$ in degrees between the ground truth direction and the predicted direction of sources as the source localization metric \cite{adavanne2018direction}. Lower $DOA_{\text{error}}$ indicates better performance. We apply the DOANet \cite{adavanne2018direction} as the baseline system. The evaluation metric is the averaged DOA error over frames:
\begin{equation} \label{eq:doa_metric}
E_{\text{DOA}} = \frac{1}{NI} \sum_{n=1}^{N}\sum_{i=1}^{I} | \hat{\Omega}_{n, i}^{(\text{DOA})} - \Omega_{n, i}^{(\text{DOA})} |
\end{equation}
\noindent where $ \hat{\Omega}_{n, i}^{(\text{DOA})} $ and $ \Omega_{n, i}^{(\text{DOA})} $ are the estimated and the ground truth DOA in the $n$-th frame, respectively. The estimated DOA $ \hat{\Omega}_{n, i}^{(\text{DOA})} $ is calculated simply by selecting directions with local maximum values on the estimated sound field. Table \ref{tab:result_doa} shows that all of NeSD-DNN, NeSD-TDNN, and NeSD-GRU systems outperform the DOANet. The NeSD-TDNN system achieves a DOA error of 0.91 degrees compared with the DOANet of 6.69 degress in one source speech localization. The DOA error increases with the number of sources in all systems. Both the NeSD-TDNN and the NeSD-GRU systems achieve similar results and outperform the NeSD-DNN system, indicating that the temporal information of sources are important for source localization.

\begin{table}
  \caption{DOA results (degree)}
  \vspace{6pt}
  \label{tab:result_doa}
  \centering
  \resizebox{\textwidth}{!}{%
  \begin{tabular}{lccccccccccccc}
    \toprule
    & & \multicolumn{4}{c}{\textbf{\textsc{Speech}}} & \multicolumn{4}{c}{\textbf{\textsc{Music}}} & \multicolumn{4}{c}{\textbf{\textsc{Sound event}}} \\
	\cmidrule(lr){3-6} \cmidrule(lr){7-10} \cmidrule(lr){11-14}
    & mic. & 1 src & 2 src & 4 src & 8 src & 1 src & 2 src & 4 src & 8 src & 1 src & 2 src & 4 src & 8 src \\
    \midrule
    DOANet \cite{adavanne2018direction} & 4 & 6.69 & 9.40 & 19.36 & 22.11 & 7.23 & 24.52 & 21.71 & 26.92 & 11.39 & 30.96 & 31.37 & 22.40 \\
    NeSD-DNN & 4 & 1.31 & 6.09 & 12.52 & 17.33 & 2.03 & 7.50 & 14.83 & 19.70 & 2.64 & 13.36 & 22.21 & 23.48 \\
    NeSD-TDNN & 4 & 0.91 & 1.93 & 4.22 & 8.49 & 1.51 & 3.70 & 8.29 & 14.44 & 1.57 & 6.50 & 16.05 & 20.68 \\
    NeSD-GRU & 4 & 1.01 & 1.70 & 4.12 & 7.46 & 1.52 & 7.66 & 13.04 & 18.21 & 1.48 & 11.69 & 22.56 & 23.73 \\
    \bottomrule
\end{tabular}}
\end{table}

\section{Conclusion}\label{section:conclusion}
We propose a learning-based neural sound field decomposition (NeSD) framework to address the sound field decomposition problem with limited number of microphones. NeSD allow sound field decomposition with fine sound direction resolution. The inputs of a NeSD system include microphone signals, microphone positions, and queried drections. The decomposed sound field can be calcualted by input arbitrary query directions to the trained NeSD system. The NeSD system can be used to predict what, where, and when are sound sources by using a unified framework. We show the the NeSD method outperforms Ambisonic and DOANet systems on wideband and mutiple sources speech, music, and sound events datasets. In future, we will explore NeSD for downstream tasks, including source localization, multiple source separation, moving object detection, and sound field reproduction tasks.
%%%%%%%%%%%%%%%%%%%%%%%%%%%%%%%%%%%%%%%%%%%%%%%%%%%%%%%%%%%%

\bibliography{ref}
\bibliographystyle{nips}

\end{document}